\documentclass[aps,pra,reprint,superscriptaddress,noeprint]{revtex4-2} % two column
\usepackage[normalem]{ulem}
\usepackage{amssymb}
\usepackage{amsmath}
\usepackage{wasysym}
\usepackage{graphicx}
\usepackage{hyperref}
\usepackage[dvipsnames]{xcolor}
\usepackage{bm}  % bold in math
\usepackage{orcidlink}
\usepackage{physics2}
\usephysicsmodule{ab}  
\usepackage{siunitx}  % qty command
\DeclareSIUnit\gauss{G}
\DeclareSIUnit\bohr{\ensuremath{a_0}}
\usepackage{csquotes}
\usepackage{braket}
\usepackage{placeins}  % \FloatBarrier before the supplement

\definecolor{jadcolor}{RGB}{33.15,173.4,237.15}

\begin{document}
\title{Observation of far-from-equilibrium scaling in the transient dynamics of 2D quantum magnets}

\author{Fabio~Bensch${}^{\orcidlink{0009-0000-1051-3297}}$}
\affiliation{Physikalisches Institut, Eberhard Karls Universit\"at T\"ubingen, Auf der Morgenstelle 14, 72076 T\"ubingen, Germany}
\author{Umberto Borla${}^{\orcidlink{0000-0002-4224-5335}}$}
\affiliation{Max Planck Institute of Quantum Optics, 85748 Garching, Germany}
\affiliation{Department of Physics and Arnold Sommerfeld Center for Theoretical Physics (ASC), Ludwig Maximilian University of Munich, 80333 Munich, Germany}
\affiliation{Munich Center for Quantum Science and Technology (MCQST), 80799 Munich, Germany}
\author{Federico~Balducci${}^{\orcidlink{0000-0002-4798-6386}}$}
\affiliation{Max Planck Institute for the Physics of Complex Systems, N\"othnitzer Str.\ 38, 01187 Dresden, Germany}
\author{Philip~Osterholz${}^{\orcidlink{0009-0007-3444-3183}}$}
\author{Shuanghong~Tang${}^{\orcidlink{0000-0002-1878-0311}}$}
\author{Silpa~Baburaj~Sheela${}^{\orcidlink{0000-0003-4415-6288}}$}
\affiliation{Physikalisches Institut, Eberhard Karls Universit\"at T\"ubingen, Auf der Morgenstelle 14, 72076 T\"ubingen, Germany}
\author{Anushya~Chandran${}^{\orcidlink{0000-0002-2046-1379}}$}
\affiliation{Department of Physics, Boston University, Boston, Massachusetts 02215, USA}
\author{Roderich~Moessner${}^{\orcidlink{0000-0002-6415-4784}}$}
\affiliation{Max Planck Institute for the Physics of Complex Systems, N\"othnitzer Str.\ 38, 01187 Dresden, Germany}
\author{Jad~C.~Halimeh${}^{\orcidlink{0000-0002-0659-7990}}$}
\affiliation{Department of Physics and Arnold Sommerfeld Center for Theoretical Physics (ASC), Ludwig Maximilian University of Munich, 80333 Munich, Germany}
\affiliation{Max Planck Institute of Quantum Optics, 85748 Garching, Germany}
\affiliation{Munich Center for Quantum Science and Technology (MCQST), 80799 Munich, Germany}
\affiliation{Department of Physics, College of Science, Kyung Hee University, Seoul 02447, Republic of Korea}

\author{Christian~Gro\ss${}^{\orcidlink{0000-0003-2292-5234}}$}
\email{christian.gross@uni-tuebingen.de}
\affiliation{Physikalisches Institut and Center for Integrated Quantum Science and Technology, Eberhard Karls Universit\"at T\"ubingen, 72076 T\"ubingen, Germany}

\date{\today}

\begin{abstract}
	The transient regime of far-from-equilibrium quantum many-body dynamics lacks the established organizing principles that universality and scaling provide in equilibrium.
	It is least understood for two-dimensional short-range interacting systems, where mean-field arguments are not expected to hold, controlled theoretical descriptions are few, and fluctuations are strong.
	Here we investigate the quench dynamics of the transverse-field Ising model using programmable Rydberg-atom arrays realizing honeycomb, square, kagome, and triangular lattices.
	Starting from a fully magnetized state, we observe a pronounced softening of the dominant collective magnetization oscillation accompanied by a maximum in the damping rate, signaling a crossover between interaction- and field-dominated transient dynamics.
	Even though the microscopic lattice geometries are different, both the oscillation frequencies and the damping rates collapse onto common curves under rescaling by the coordination-number-weighted interaction strength.
	We present a theoretical description which accounts for the collapse of both quantities and elucidates the nature of the processes responsible for the damping.
	The observed failure of the discrete truncated Wigner approximation to predict the damping for strong interactions shows that classical evolution of initial quantum noise is insufficient, while tree-tensor-network simulations reproduce the dynamics accurately.
	These results reveal a robust scaling regime governing the transient dynamics of short-range interacting two-dimensional quantum magnets.
	The fact that the dominant transient dynamics can be captured by a simple collective description despite the presence of fluctuations and nonlinearities characteristic of low dimensionality and small spin is an important insight in the quest for organizing principles in far-from-equilibrium quantum matter.
\end{abstract}

\maketitle

% General introduction - without universality and prethermal
The transient regime of far-from-equilibrium dynamics in higher-dimensional quantum many-body systems is notoriously difficult to describe~\cite{polkovnikov2011-jch,eisert2015-jch,gogolin2016,dalessio2016,ueda2020}. Although controlled numerical methods can capture short-time dynamics, the rapid growth of entanglement severely limits the accessible timescales, particularly in two spatial dimensions~\cite{orus2014,cirac2021}. At later times, hydrodynamic and other effective descriptions become available when the dynamics is governed by conserved quantities, slow modes, or low-energy excitations~\cite{mori2018-jch}.
The intervening transient regime is therefore both theoretically challenging and physically important: it contains the coherent response generated directly by the quench, determines the initial conditions for subsequent relaxation, and constitutes the principal time window accessible to present-day quantum simulators and quantum processors~\cite{monroe2021-cg,browaeys2020-cg,andersen2025-cg}.
Whether simple organizing principles can nevertheless emerge in this regime remains an open question.

Quantum quenches have been extensively studied theoretically in fully connected and other mean-field models, where they can exhibit an order-parameter dynamical phase transition: the long-time or prethermal behavior of the magnetization changes nonanalytically at a dynamical critical point~\cite{barankov2006-cg,yuzbashyan2006-cg,sciolla2010-cg,sciolla2011-jch,homrighausen2017-jch,lang2018-jch}.
Within the semiclassical description, the dynamical critical point corresponds to a separatrix of the collective-spin motion: the oscillation period diverges, and the collective magnetization mode consequently softens~\cite{sciolla2011-jch,Lerose2019Impact,marino2022-cg}. 
Signatures of this phenomenology have been observed in platforms with long-range or effectively all-to-all interactions~\cite{zibold2010,klinder2015-cg,trenkwalder2016,zhang2017-cg,jurcevic2017-cg,smale2019-cg,yang2019-cg,muniz2020-cg,tian2020-cg,xu2020-cg,young2024-cga,schuckert2025-cg,champion2025-cg,natale2026-cg,bullock2026-cg,austin-harris2026-cg}.

For short-range interacting systems, spatial fluctuations, quasiparticle propagation, and eventual thermalization invalidate the exact collective-spin description and can replace a sharp asymptotic dynamical transition by a finite-time crossover~\cite{dag2021-cg}.
Numerical studies of the short-range two-dimensional transverse-field Ising model have nevertheless found signatures of dynamical criticality on accessible timescales~\cite{hashizume2020-jch,hashizume2022-cg,balducci2026-cg}.
What remains largely unexplored is whether the transient response of such systems obeys simple scaling principles across microscopically distinct two-dimensional lattice geometries.
More broadly, can transient quantum dynamics exhibit organizing principles analogous to the powerful scaling structures in equilibrium statistical mechanics~\cite{wilson1971-jch,fisher1974-jch,Zvyagin20162016-jch,berges2021,mikheev2023}?

Here we experimentally, numerically, and analytically investigate far-from-equilibrium transients following global quenches of the short-range transverse-field Ising model in two spatial dimensions.
By comparing honeycomb, square, kagome, and triangular lattices, we uncover a single-parameter scaling regime governing their dominant collective response.
The dynamics is dominated by decaying magnetization oscillations whose frequency exhibits a pronounced mode softening as the quenched transverse field is varied, an effect that has also been observed in antiferromagnetic two-dimensional Rydberg arrays subject to a staggered field~\cite{manovitz2025-cg, balducci2026-cg}.
We show that this mode softening follows from an effective mean-field description of the collective magnetization dynamics and establish that this description remains applicable to short-range interacting two-dimensional magnets on transient timescales.
After rescaling by the coordination-number-weighted interaction strength, the field-dependent oscillation frequencies collapse across all four lattices.
The collective response is furthermore robust against vacancies and the strong local longitudinal fields present at edges and corners, with bulk, edge, and corner spins sharing a common dominant oscillation frequency.
Even more remarkably, the fluctuation-induced damping rates, which lie beyond the mean-field description of the uniform magnetization, collapse under the same rescaling despite the presence of correlated quantum fluctuations, which we directly measure.
We present a semiclassical and perturbative treatment of spatial fluctuations atop the collective magnetization oscillation, which predicts the shape of the damping curve and its collapse under rescaling, despite being quantitatively inaccurate for spin-1/2 systems.
Comparisons of the measured data with semiclassical and tensor-network simulations reveal a hierarchy of descriptions: mean-field theory captures the collective frequency, large-spin semiclassics predicts the presence of a scaling collapse of the damping, the discrete truncated Wigner approximation predicts the damping --- quantitatively correct 
only in the field-dominated regime --- and tree-tensor-network simulations reproduce the microscopic correlation dynamics accurately.

\begin{figure}[t]
	\centering \includegraphics{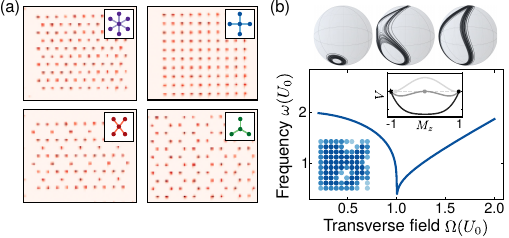}
	\caption{
		Illustration of the experimental system and MF interpretation.
		\textbf{(a)} Average fluorescence images (red indicates more photon counts per camera pixel) for the four different lattice geometries studied here.
		The insets highlight the different local connectivity of the lattices.
		\textbf{(b)} MF dynamics for the square lattice with nearest-neighbor interactions. At the top, trajectories of the mean magnetization vector are illustrated for transverse fields of $\Omega/U_0 = \numlist{0.5;1.0;1.5}$ and zero longitudinal field.
		The individual trajectories correspond to different initial states.
		The plot below shows the predicted mode softening of the post-quench oscillation frequency as a function of the transverse field $\Omega/U_0$ for the example of a square lattice and vanishing longitudinal field.
		The lower left inset illustrates the local longitudinal fields for the square lattice example, where lighter color encodes stronger fields.
		The upper inset shows the effective MF potential $V$ for the same three values of $\Omega/U_0$ as used for the trajectory illustrations (darker for larger $\Omega/U_0$).
		The energies of the individual curves are offset such that the initial state lies at zero energy for each of them.
		The star indicates the initial state at $M_z=-1$ and zero energy (dashed gray line), the points indicate the classical turning points.
	}
	\label{fig:1}
\end{figure}

% Minimal intro of the physical system
Our experiments are performed with potassium-39 atoms held in two-dimensional arrays of optical tweezers, a fully controlled platform to explore the far-from-equilibrium quench response of transverse-field quantum Ising models~\cite{lienhard2018-cg,guardado-sanchez2018-cg,keesling2019-cg,bornet2023-cg,manovitz2025-cg,darbha2025-cg}.
The platform, including all experimental implementation details, is described in Ref.~\cite{osterholz2025}.
The pseudospin-$1/2$ is encoded in an atomic ground state $\ket{\downarrow}$ and a Rydberg state $\ket{\uparrow}$, whose specific choice we detail later.
Atoms in the Rydberg state interact via the van der Waals potential $U_{ij} = U_0/r_{ij}^6$, where $U_0$ is the nearest-neighbor interaction strength, and $r_{ij}$ measures the distance between atoms at sites $i$ and $j$ in units of the nearest-neighbor distance $d_0$.
In two dimensions, such an interaction exhibits long-wavelength properties characteristic of strictly short-ranged systems, in contrast to the long-range interacting systems for which mean-field (MF) descriptions are well established~\cite{defenu2023}.
Since the tweezer positions are freely programmable, the lattice geometry can be configured at will, and we realize arrays of different effective coordination numbers $\widetilde{\mathcal{N}}_c$: approximately $\num{6.4}$ for the triangular, $\num{4.7}$ for the square, $\num{4.3}$ for the kagome, and $\num{3.3}$ for the honeycomb lattice.
The fractional values arise from the longer-ranged tail of the interactions, which affects the lattice geometries differently and which we account for by weighting each neighbor of a site $i_0$ with its interaction strength, $\widetilde{\mathcal{N}}_c = \sum_{i\neq i_0} r_{i,i_0}^{-6}$.
The four geometries studied here are shown in \autoref{fig:1}a.

% Hamiltonian
The realized Rydberg Ising Hamiltonian is~\cite{osterholz2025}
\begin{equation}
	\hat{H}/\hbar = \frac{1}{8} \sum_{i \neq j} U_{ij} \hat{Z}_i\hat{Z}_j - \frac{h_z}{2} \sum_i \hat{Z}_i + \frac{\Omega}{2} \sum_i \hat{X}_i,
	\label{eq.hamiltonian}
\end{equation}
where $\hbar$ is the reduced Planck constant.
$\hat{X}_i$, $\hat{Z}_i$ are Pauli operators at site $i$, and $\hat{Z}_i$ is related to the projector onto the Rydberg state $\hat{R}_i$ by the relation $\hat{R}_i =(\hat{Z}_i + \hat{\mathbb{I}}_i)/2$ with the identity operator $\hat{\mathbb{I}}_i$.
Since $U_0>0$, the Ising coupling is antiferromagnetic.
Preparing the fully magnetized state, however, places us in the high-energy sector of the spectrum, where the dynamics is that of a ferromagnet and the frustration of the triangular and kagome lattices plays no role.
In an infinite system, this Hamiltonian in standard Ising form arises from the Rydberg Hamiltonian by the introduction of an effective longitudinal field $h_z = \Delta-\widetilde{\mathcal{N}}_c U_0/2$ and the transverse field $\Omega$.
Here, $\Delta$ and $\Omega$ are the detuning and the Rabi frequency of the laser that uniformly couples the ground and Rydberg states~\cite{osterholz2025}.
Sites with fewer interaction partners, such as those at the edges and corners of a finite array or next to an atom missing after the rearrangement, experience a local longitudinal offset field $h_{\mathrm{loc},i} = \frac{1}{2}\sum_{j} \varnothing_{j} U_{ij}$, where $\varnothing_{j}$ is one for empty sites $j$ and zero otherwise.
For van der Waals interactions, both the couplings and the shifts in the global and local longitudinal field are dominated by the nearest neighbors, and we can often approximate $h_z \approx \Delta-{\mathcal{N}}_c U_0/2$ and $h_{\mathrm{loc},i} \approx \mathcal{N}_{\mathrm{vac},i} U_0/2$, where $\mathcal{N}_c$ is the number of nearest neighbors and $\mathcal{N}_{\mathrm{vac},i}$ counts missing neighbors.

% Initial state and long wavelength argument
In all experiments reported here, we initialize the system in the fully magnetized ferromagnetic state with all spins down $\ket{\psi_0} = \ket{\downarrow\downarrow\ldots\downarrow}$, and we fix the laser detuning to $\Delta = \mathcal{N}_c U_0/2$ such that the effective longitudinal field generated by the $\mathcal{N}_c$ nearest neighbors is canceled.
Exact cancellation in the bulk would instead require $\Delta = \widetilde{\mathcal{N}}_c U_0/2$, which also accounts for the interaction tail.
However, realizing exactly zero longitudinal field is experimentally challenging due to the presence of edges and corners, as well as the random distribution of missing atoms after the rearrangement procedure~\cite{osterholz2025}.
Averaging over these effects, the per-atom longitudinal field in our experiments is $\bar{h}_z/U_0 = \numlist{0.3;-0.15;0.19;0.06}$ for the triangular, square, kagome, and honeycomb lattices.
Importantly, the driving laser field is significantly detuned with respect to the isolated spin-flip resonance $\Delta=0$, thereby suppressing the appearance of small-scale domains, in particular in the range $U_0 > \Omega$.
This setting is expected to restrict the transient dynamics to the long-wavelength sector, with relatively small local fluctuations.
One may therefore expect an MF description to capture the global magnetization dynamics on short timescales, even though the underlying microscopic model is short-ranged and strongly interacting.

% Short mean field intro
Within the MF approximation, the two-dimensional transverse-field Ising Hamiltonian reduces to a single collective spin, described by the magnetization per site $\vec{M}$ averaged over the lattice, which precesses according to $\dot{\vec{M}} = \vec{M} \times \vec{B}$ in the field $\vec{B} = (-\Omega,\, 0,\, h_z-U_0 \widetilde{\mathcal{N}}_c M_z/2)$.
Since both the MF energy and the spin length are conserved, the dynamics can be recast as that of a one-dimensional particle in an effective quartic potential $V(M_z)$ whose shape is determined, for small $h_z$, by the ratio $\Omega/(U_0\widetilde{\mathcal{N}}_c)$ (see \autoref{sec:mf}).
For weak transverse fields, the quench energy of the initially fully magnetized state confines the magnetization to a single well of a double-well potential, such that the post-quench time-averaged magnetization remains non-zero.
Above the critical value $\Omega_{\mathrm{crit}} = U_0 \widetilde{\mathcal{N}}_c / 4$, at which the quench energy exactly matches the energy of the maximum at $M_z=0$, the time-averaged magnetization is zero and the classical turning points have equal and opposite $M_z$ values.
Right at $\Omega_{\mathrm{crit}}$ the magnetization comes to rest at $M_z=0$ and the oscillation frequency vanishes.
MF theory thus predicts a minimum of the oscillation frequency at the critical transverse field (\autoref{fig:1}b)~\cite{balducci2026-cg}, a phenomenon we refer to as \textit{mode softening}, which provides an observable to measure the dynamical critical point.
This picture applies for small longitudinal fields $h_z$.
Below we show experimentally that the MF description is indeed valid for the magnetization dynamics.

\begin{figure}[t]
	\centering \includegraphics{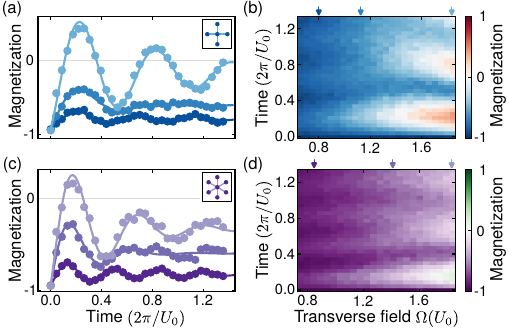}
	\caption{
		Post-quench magnetization dynamics of $M_z$ in square and triangular lattices, row-wise as identified by the pictogram-insets in (a) and (c).
		\textbf{(a)} Square lattice magnetization traces for $\Omega/U_0 = \numlist{0.8;1.3;1.8}$, from darker to lighter color.
		The fits are shown as solid lines and the zero crossing of the magnetization is indicated by the light gray line.
		\textbf{(b)},\textbf{(d)} Color plots of the square (b) and triangular (d) lattice dynamics.
		The arrows on top indicate the three values of $\Omega/U_0$ shown in (a) and (c).
		\textbf{(c)} Triangular lattice analog of (a) for $\Omega/U_0 = \numlist{0.9;1.4;1.8}$.
		In both cases the three values lie in the interaction-dominated regime, near the crossover, and in the field-dominated regime.
		Statistical error bars of one standard error of the mean are smaller than the marker size.
		See~\autoref{fig:si1} for the same data for the kagome and honeycomb lattices.
	}
	\label{fig:2}
\end{figure}

% Short TWA and damping model
By construction, the deterministic MF description does not capture fluctuations that eventually damp the magnetization oscillations.
In order to include damping we extend the theoretical description perturbatively by an analytically tractable approximate treatment of the spin waves (see \autoref{sec:semiclassical}).
In this model, long-wavelength modes grow exponentially around the point where the magnetization vanishes, $M_z=0$, with a 
growth rate that depends principally on the lattice coordination. 
The damping is dominated by this growth rate, and hence displays scaling behavior; therefore, the model predicts a universal form of the damping rate . 

In order to connect with the spin-1/2 experimental setting, we extend the semiclassical treatment via a discrete truncated Wigner approximation (DTWA) for the square lattice model~\cite{schachenmayer2015-jch}.
Leaving the details to \autoref{sec:DTWA}, the DTWA shares with the semiclassical treatment the factorized ansatz for phase-space operators, representing the many-body density matrix. It further tries to capture quantum fluctuations through a discrete-Wigner-function sampling of the initial state~\cite{Wootters1987Wigner,wootters2003picturing}. Each sample is then evolved classically. 
The classical evolution is generated by the classical limit of the Hamiltonian in \autoref{eq.hamiltonian}, and averaging over the trajectories yields expectation values of global observables that are qualitatively correct in the field-dominated regime~\cite{Vovrosh:2026aa}.

We complement the experiments with tree-tensor-network (TTN) \cite{tagliacozzo2009-jch,murg2010-jch} simulations, a tensor-network approach that efficiently captures quantum correlations in two-dimensional systems beyond MF and semiclassical descriptions. Because of the high computational cost, we restrict the TTN simulations to the square lattice, which serves as a representative benchmark geometry.

% FIGURE 2
% Measurements and extraction of the frequency and the damping rate
We first aim to study the validity of the MF description of the dynamics and to observe the predicted mode softening.
The atomic ground state $\ket{\downarrow}$ is $4\mathrm{S}_{1/2} \, \ket{F=2,m_F=2}$ and the Rydberg state $\ket{\uparrow}$ is the $62\mathrm{S}_{1/2} \, \ket{m_J=1/2}$ state for all measurements except for the large $15\times15$ square array, for which we use the $51\mathrm{S}_{1/2}\,\ket{m_J=1/2}$ state (see \autoref{sec:experiment}).
For the $62\mathrm{S}_{1/2}$ state we set a nearest-neighbor distance of $d_0 = \qty{7}{\micro\meter}$, resulting in $U_0=2\pi \times \qty{1.7}{\mega\hertz}$, while for the $51\mathrm{S}_{1/2}$ state we set $d_0 = \qty{5}{\micro\meter}$ ($U_0 = 2\pi \times \qty{1.0}{\mega\hertz}$).
Starting with $\ket{\psi_0}$, we suddenly quench the transverse field from zero to a variable final value $\Omega$ and measure the real-time dynamics of the magnetization $M_z = \langle \hat{M}_z \rangle$, where $\hat{M}_z = \frac{1}{N}\sum_i \hat{Z}_i$.
\autoref{fig:2} shows the measured time evolution for the square and the triangular geometry.
The magnetization dynamics changes qualitatively when going from the small-$\Omega$, interaction-dominated to the large-$\Omega$, field-dominated regime.
In the former, the magnetization keeps its initial direction on average and shows small and fast oscillations.
In the latter, it performs large-amplitude oscillations and swings to the opposite side.
In the intermediate regime, the dynamics damps out quickly.
The main features of the magnetization traces are captured by an exponentially damped harmonic fit $M_z = A \exp(-\gamma t) \cos(\omega t + \varphi)+C$, from which we extract the oscillation frequency $\omega$ and the damping rate $\gamma$.
TTN simulations, detailed in \autoref{sec:ttn}, reproduce our experimental observations quantitatively.

% FIGURE 3

\begin{figure}[t]
	\centering \includegraphics{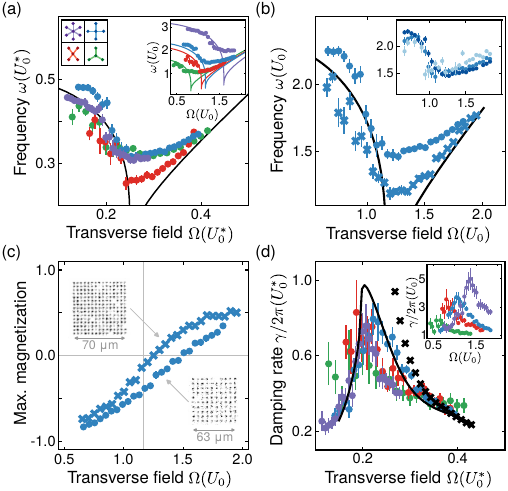}
	\caption{
		Mode softening and damping.
		Colors indicate the different lattices as shown by the upper right inset in (a).
		\textbf{(a)} Magnetization oscillation frequency $\omega$ for the four lattice geometries.
		Solid lines are the MF predictions at zero longitudinal field.
		The inset shows the unscaled data, while in the main panel both axes are scaled by the coordination-number-weighted interaction strength $U_0^*$.
		\textbf{(b)}, \textbf{(c)} Finite-size effects: comparison of the $10\times10$ (circles) and $15\times15$ (crosses) square lattices, for the oscillation frequency (b) and the maximum magnetization swing (c).
		The mode softening is more pronounced for the larger array.
		The inset of (b) compares the measured frequencies post-selected to only bulk, edge, or corner spins (increasing lightness).
		The insets of (c) show two single shots for the two array sizes and the light gray lines indicate the zero magnetization and the MF prediction for the dynamical critical point.
		The maximum magnetization swing is extracted from the time traces by taking the average of the three largest magnetization values.
		\textbf{(d)} Damping rate $\gamma$ in original units (inset) and scaled by the coordination-number-weighted interaction strength $U_0^*$ (main panel).
		The black line is the TTN simulation for the square lattice including the experimental imperfections; the black crosses are the DTWA prediction, both fitted by the same exponentially damped model as the data.
		The DTWA fits fail in the interaction-dominated regime, where no data points are shown.
		Error bars represent the uncertainties in the fit parameters in all panels but (c), where they are one standard error of the mean. See \autoref{fig:si2} for the results of the other fit parameters.
	}
	\label{fig:3}
\end{figure}
% Mode softening
Plotting the fitted frequencies as a function of the transverse field reveals the mode softening in the intermediate-field regime (inset of \autoref{fig:3}a).
The observed behavior is consistently reproduced by the MF description for all four lattice geometries, up to a broadening of the feature observed in the experiment.
Scaling the transverse field and the collective frequency by the coordination-number-weighted interaction strength $U_0^* = \widetilde{\mathcal{N}}_c U_0$ leads to a collapse of both data and theory (\autoref{fig:3}a).
The position of the frequency minimum collapses well around $\Omega/U_0^* \approx 0.25$, in good agreement with the MF value $\Omega_{\mathrm{crit}}/U_0^* = 1/4$.
The broadening of the experimental data can in large part be attributed to the finite system size, as a direct experimental comparison of the $10\times10$ square array data with those from an analogous experiment in a $15\times15$ setting reveals (\autoref{fig:3}b).
The mode softening has been discussed as a signature of the dynamical critical point of the quenched dynamics, which in our finite, short-range system appears as a crossover; for brevity, we refer to the location of this crossover as the dynamical critical point throughout.
While dynamical critical points are typically explored in a zero-longitudinal-field setting, our data suggest a generalization of the concept to settings with a finite density of sites subject to local longitudinal fields.

% Effects of local fields
Finite local longitudinal fields also broaden the mode softening (see~\autoref{fig:si3}a).
They are, however, not what distinguishes the two array sizes: the fraction of sites with all nearest neighbors present increases only slightly, from $\num{0.42}$ to $\num{0.49}$, for the larger square array.
To study the influence of the local fields, we selectively analyze the magnetization dynamics for sites in the bulk, along the edges, and on the corners of the $10\times10$ square array (inset of \autoref{fig:3}b).
The edge and corner fields are stronger than residual bulk fields by $U_0/2$ and $U_0$, respectively.
Despite this strong local bias, we observe a uniform frequency of the magnetization oscillation, that is, the different populations lock in frequency.
This aspect is explained by a two-population extension of the MF model (see \autoref{sec:mf}), in which the interaction between the populations is what locks them.
The collective mode is thus robust against the strong local fields and the finite density of vacancies present in the array, a robustness that is not evident from the MF description itself, in which all sites are equivalent.

% Max magnetization swing
Another signature of the dynamical critical point is the vanishing temporal average of the magnetization at the critical point.
However, this is not a robust observable if the observation time is limited.
An alternative observable is the maximum magnetization swing, the maximum of the magnetization reached in the oscillation, a proxy for the classical turning point of the MF dynamics. 
Its zero crossing signals the location of the dynamical critical point and can be extracted from transient dynamics.
In \autoref{fig:3}c we show the maximum magnetization swing reached during the dynamics for different $\Omega$.
Comparing the smaller and the larger square array reveals a consistent reduction of the amplitude in the smaller array and an approach to the field-free MF prediction for the larger array.

% Damping
Having established that the coherent collective frequency is controlled by the single scale $U_0^*$ and remains robust against strong local inhomogeneities, we now ask whether the fluctuation-induced damping exhibits the same collective organization.
The damping rate $\gamma$ extracted from the fit to the time-evolution data is shown in \autoref{fig:3}d.
For all lattice geometries we observe a maximum that coincides with the minimum in the oscillation frequency.
With increasing effective coordination number $\widetilde{\mathcal{N}}_c$, the variation of the damping becomes more pronounced and the maximum becomes sharper.
Rescaling the transverse field and damping rate with the coordination-number-weighted interaction strength $U_0^*$ again leads to a data collapse, showing that the damping, too, is approximately independent of the geometry.
Our measurements therefore suggest that the same interaction scale $U_0^*$ controls two distinct physical processes: the coherent collective precession captured by MF theory and its damping.
Fitting the TTN data in the same way as the experimental traces results in close agreement between simulation and experiment.
We explore the DTWA as one minimal model for the damping of the magnetization oscillations.
It reproduces the increase of the damping on approaching the crossover from the field-dominated side reasonably well, but overestimates the damping in the interaction-dominated regime, so much, that the fit to a damped exponential is unreliable.

The semiclassical spin-wave extension to the MF provides analytical insight and an intuitive physical picture of the observed scaling behavior.
We have analyzed the temporal growth of fluctuations, which reproduces the scaling collapse observed experimentally (see \autoref{eq:Hill}): the growth of fluctuations is dominated by the part of the MF trajectory near vanishing magnetization $M_z$, where its  
rate depends on the lattice structure principally through the (effective) coordination number. Note that this exponential growth rate is independent of the strength of the initial fluctuations.
We validate this interpretation by numerically solving the semiclassical equations of motion, confirming the scaling behavior in a wide range of initial fluctuation amplitudes.

\begin{figure}[t]
	\centering \includegraphics{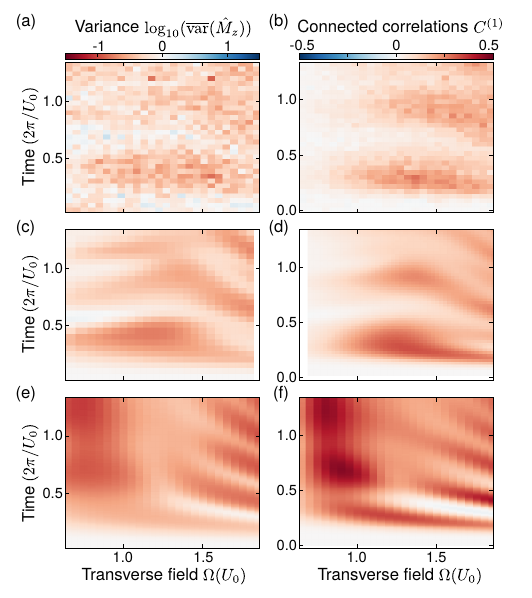}
	\caption{
		Fluctuations and correlations in the $10\times10$ square array: the TTN simulations reproduce the measurements, while the DTWA breaks down in the interaction-dominated regime.
		The base-10 logarithm of the normalized variance of the magnetization, $\log_{10}(\overline{\mathrm{var}}(\hat{M}_z))$, is shown in the left column such that zero corresponds to the variance of non-interacting spins.
		\textbf{(a)} is the experimental data, \textbf{(c)} the TTN simulation, and \textbf{(e)} the DTWA result.
		In the right column we show the nearest-neighbor connected correlator $C^{(1)}$ in the same order as before, that is,
		\textbf{(b)}, the experimental data, \textbf{(d)}, the TTN simulation, and \textbf{(f)}, the DTWA result.
	}
	\label{fig:4}
\end{figure}

% FIGURE 4
% Fluctuations of the magnetization and two-point correlators
To shed light on the dynamics of collective fluctuations and correlations, we analyze the fluctuations of the magnetization, globally and locally, for the $10\times10$ square lattice setting, for which our measurements have the best statistics.
Overall, the fluctuation analysis reveals excellent agreement between the TTN simulations and the experiment (see \autoref{fig:4} and, for a more detailed comparison,~\autoref{fig:si4}).
We measure the global fluctuation by the normalized variance of the magnetization $\overline{\mathrm{var}}(\hat{M}_z) = N( \langle \hat{M}_z^2\rangle - \langle \hat{M}_z\rangle^2)/ (1-\langle \hat{M}_z\rangle^2 )$, where the normalization is with respect to the variance of a binomial distribution, which is expected for a non-interacting ensemble of $N$ spins.
On average, we find an enhanced variance, supporting the interpretation of the observed collective scaling in the damping of the magnetization oscillations.
As a representative local observable, we choose the nearest-neighbor connected correlator $C^{(1)} = \langle \langle \hat{Z}_i \hat{Z}_j \rangle - \langle \hat{Z}_i \rangle \langle \hat{Z}_j \rangle \rangle$, where the inner averaging is over different shots and the outer is over all statistically independent nearest-neighbor bonds.
Both quantities, $\overline{\mathrm{var}}(\hat{M}_z)$ and $C^{(1)}$, show similar qualitative behavior in their dynamics, with no distinctive feature at the dynamical critical point other than a broad transient maximum as a function of the transverse field.
In particular, they show no indication of enhanced damping of their dynamics near the dynamical critical point, where especially $C^{(1)}$ shows indications of longer-lived oscillations (see also~\autoref{fig:si4}).
This suggests that the signatures of dynamical criticality most prominently affect global observables, in agreement with our collective interpretation of the effect.
The most prominent feature of the comparison, however, is the failure of the DTWA to describe the experiment toward the interaction-dominated regime.
While the oscillating dynamics is at least qualitatively reproduced in the field-dominated regime, the DTWA predicts starkly different dynamics at the dynamical critical point and beyond, with strong fluctuations and correlations in regions where both experiment and TTN find them to be weak.

% Conclusion
We have experimentally investigated the transient quench dynamics of the two-dimensional transverse-field Ising model realized with programmable Rydberg-atom arrays on honeycomb, square, kagome, and triangular lattices.
Despite their distinct microscopic connectivities, all four geometries exhibit the same collective response, controlled by the coordination-number-weighted interaction strength $U_0^*$: a pronounced softening of the dominant magnetization oscillation frequency accompanied by a maximum in the damping rate.
Upon rescaling by $U_0^*$, both quantities collapse onto common curves, demonstrating that the dominant collective response is largely controlled by this single interaction scale rather than by the microscopic lattice geometry.

While the collapse of the oscillation frequency immediately follows from the MF description, the damping involves processes beyond it; our measurements show that $U_0^*$ also sets the scale at which the collective mode damps, which is backed up by our further theoretical modeling.
This physical insight places the damping we observe in the realm of non-viscous friction \cite{kimball1927-cg,Zener1948}: it is independent of the speed at which the processes take place, as well as linear in the sense of being independent of the strength of the fluctuations, in the tradition of Newton's treatment of what is now known as the coefficient of restitution \cite{Newton1687Principia}. 
These findings are based on the experimental observation of the scaling collapse: the former explicitly, the latter implicitly, as we do not actively manipulate the relative strengths of the initial fluctuations across the different lattice geometries. 

The observed hierarchy of theoretical descriptions provides additional insight into the nature of this transient regime.
While a simple MF theory accurately captures the collective oscillation frequency and its softening despite the presence of strong quantum fluctuations and local fields, and while a spin-wave treatment can qualitatively predict the presence of a scaling collapse in the damping curves, the DTWA is needed to reproduce the fluctuation-induced damping in the field-dominated regime, and TTN simulations to describe the evolution of correlation functions.
Together these results indicate that increasingly sophisticated descriptions become necessary near the crossover and as progressively more microscopic observables are considered.
Our work identifies a remarkably robust collective response in the far-from-equilibrium dynamics of short-range interacting quantum magnets and establishes a simple organizing principle for transient nonequilibrium dynamics in two spatial dimensions.
An important open question is whether this discovered organizing principle extends to the transient response of other short-range quantum systems, including frustrated magnets and lattice gauge theories, where analogous quench protocols can be implemented on the same programmable platforms~\cite{semeghini2021,halimeh2025a,halimeh2025b}.
If so, this would point toward a genuinely universal picture for transient far-from-equilibrium dynamics.

%%%%%%%%%%%%%%%%%%%%%%%%%%%%%%%%%%%%%%
% End
%%%%%%%%%%%%%%%%%%%%%%%%%%%%%%%%%%%%%%

\bigskip
%\begin{acknowledgments}
\textbf{Acknowledgments:}
We thank Roland C.~Farrell, Lucas Katschke, and Jesse J.~Osborne for fruitful discussions. This work received funding from the Horizon Europe program HORIZON-CL4-2022-QUANTUM-02-SGA via the project 101113690 (PASQuanS2.1) and via the Horizon-MSCA-Doctoral Network QLUSTER (HORIZON-MSCA-2021-DN-01, GA 101072964), the Federal Ministry of Education and Research Germany (BMBF) via the project 13N15974, and the Deutsche Forschungsgemeinschaft within the research units FOR 5413 (Grant No. 465199066) and Cluster of Excellence
ct.qmat (EXC 2147, Project-ID No. 390858490).
We also acknowledge funding through JST-DFG 2024: Japanese-German Joint Call for Proposals on \enquote{Quantum Technologies} (Japan-JST-DFG-ASPIRE 2024) under DFG Grant No. 554561799 and from the Alfried Krupp von Bohlen and Halbach Foundation.
U.B.~and J.C.H.~acknowledge funding by the Max Planck Society, the Deutsche Forschungsgemeinschaft (DFG, German Research Foundation) under Germany’s Excellence Strategy – EXC-2111 – 390814868, and the European Research Council (ERC) under the European Union’s Horizon Europe research and innovation program (Grant Agreement No.~101165667)—ERC Starting Grant QuSiGauge. A.C. also acknowledges support by the Air Force Office of Scientific Research (AFOSR) under Grant No. FA9550-24-1-0121.
The numerical simulations were partly performed on the MPIPKS HPC cluster.
%\end{acknowledgments}

% \bigskip
% \textbf{Author Contributions:} All authors contributed extensively to the
% planning, data acquisition, or analysis of the results presented here. 

\bigskip
\textbf{Data availability:}
The experimental and theoretical data and evaluation scripts that support the findings of this study will be available on Zenodo.

\bigskip
\textbf{Competing interests:}
There are no competing interests to declare.

\bigskip
\textbf{Note:} During the completion of this work, we became aware of a related complementary experiment by Pasqal investigating far-from-equilibrium dynamics in two-dimensional Rydberg Ising systems \cite{PasqalExperiment}. While that work demonstrates signatures of dynamical critical behavior in a single square lattice geometry and focuses on the thermalization dynamics, the present work addresses a different question: the emergence of geometry-independent scaling across multiple two-dimensional lattice geometries and its description through a hierarchy of effective theoretical approaches ranging from mean-field theory to discrete truncated Wigner and tensor-network methods.

\bibliography{bibliography}

%%%%%%%%%%%%%%%%%%%%%%%%%%%%%%%%%%%%%%%%%%%%%%%%%%%%%%%
%%%%%%%%% Supplemental Material %%%%%%%%%%%%%%%%%%%%%%%
%%%%%%%%%%%%%%%%%%%%%%%%%%%%%%%%%%%%%%%%%%%%%%%%%%%%%%%

\FloatBarrier

\clearpage

\section*{Methods}
% \section*{Supplemental material}

% Number supplement figures and equations S1, S2, ... (the \FloatBarrier above
% flushes all main-text floats first, so the counter reset cannot renumber them)
\setcounter{figure}{0}
\renewcommand{\thefigure}{S\arabic{figure}}
\setcounter{equation}{0}
\renewcommand{\theequation}{S\arabic{equation}}

\subsection{Experiment preparation}
\label{sec:experiment}
The optical tweezer arrays are generated from a \qty{1064}{\nano\meter} single-mode fiber combined with a spatial light modulator (SLM) placed in the Fourier plane of the in-vacuum objective.
We used two different array dimensions for the square geometry, $15\times15$ and $23\times23$, together with a triangular array of $15\times16$ sites.
After rearrangement, we obtain $10\times10$ and $15\times15$ square arrays, as well as a $10\times10$ triangular array and kagome and honeycomb arrays with 72 and 88 atoms, respectively.
The triangular, kagome, and honeycomb arrays all originate from the same underlying triangular pattern: during the rearrangement procedure, atoms are removed from selected sites to realize the kagome and honeycomb structures~\cite{osterholz2025}.
Only the square arrays are based on a separate tweezer pattern.
For the $15\times16$ triangular and the $15\times15$ square tweezer patterns, the nearest-neighbor distance is $d_0 = \qty{7}{\micro\meter}$, whereas for the $23\times23$ pattern it is $d_0 = \qty{5}{\micro\meter}$.
This choice keeps the overall array size comparable across geometries, so that the Gaussian envelope of the Rydberg beams provides similar intensity homogeneity across the array.
The two Rydberg states used are $51\mathrm{S}_{1/2}\,\ket{m_J=1/2}$ for the $d_0 = \qty{5}{\micro\meter}$ arrays ($U_0 = 2\pi \times \qty{1.0}{\mega\hertz}$) and $62\mathrm{S}_{1/2}\,\ket{m_J=1/2}$ for the $d_0 = \qty{7}{\micro\meter}$ arrays ($U_0 = 2\pi \times \qty{1.7}{\mega\hertz}$).
The use of two distinct tweezer patterns explains the slight systematic mismatch between the experimental data and the MF prediction in the interaction-dominated regime: the square-array data are shifted to slightly larger frequencies, while the other three geometries match (see~\autoref{fig:3}).
We attribute this to a small difference in the lattice constant $d_0$ between the two underlying patterns.
Assuming a slightly larger interaction strength of $U_0'=2\pi\times \qty{1.75}{\mega\hertz}$ restores agreement for the square lattice, and a systematic shift in lattice spacing of only $\qty{0.5}{\percent}$ ($\qty{35}{\nano\meter}$) accounts for this $\qty{3}{\percent}$ change in $U_0$.

After a magneto-optical trap (MOT) and a compression stage, atoms are loaded into a thin \qty{1064}{\nano\meter} dipole trap with waists of \qty{7}{\micro\meter} (vertical) and \qty{130}{\micro\meter} (horizontal), from which they are subsequently loaded into the tweezers under gray molasses cooling.
We use the gray molasses beams for parity projection, after which we reach a loading probability of \qty{65}{\percent}.
We induce fluorescence on the D2 line with simultaneous cooling to detect atoms in the tweezers.
Due to the strong polarizability of the D1 and D2 excited states to the \qty{1064}{\nano\meter} light (more than six times larger than that of the ground state), achieving a survival probability above \qty{99}{\percent} during the imaging phase requires chopping the D2 light at \qty{1}{\kilo\hertz}, which is far from any typical trap frequency of the tweezers.
Another crucial aspect for achieving high-fidelity, high-survival-probability imaging is the inhomogeneity of the tweezer array.
We use the D2 fluorescence counts to feed back the amplitude of each tweezer, achieving $<\qty{1}{\percent}$ inhomogeneity.

After acquiring the first image of the array, we begin the sorting phase, in which we identify which tweezers need to be emptied and which need to be filled with an atom.
We use a crossed acousto-optic deflector (AOD) to reshuffle the atoms; a \qty{795}{\nano\meter} distributed-Bragg-reflector (DBR) laser drives the AOD tweezer.
In a single sorting run, we reach a \qty{90}{\percent} filling fraction in the target tweezers, with a loading probability of $<\qty{1}{\percent}$ from tweezers meant to be emptied.

Gray molasses cooling does not bring the atoms to a low enough temperature to neglect thermal disorder in our simulations.
We therefore use a Raman sideband cooling (RSC) procedure to cool the atoms to the vibrational ground state of the tweezer's harmonic potential.
We achieve \qty{90}{\percent} (\qty{85}{\percent}) radial (axial) ground-state probability, corresponding to an overall 3D ground-state probability of roughly \qty{70}{\percent}.
After the RSC procedure, the atoms are prepared in the $\ket{4S_{1/2},F=2, m_F=2}$ state, which in this context denotes the fully magnetized state $\ket{\psi_0} = \ket{\downarrow\downarrow\ldots\downarrow}$.

\subsection{Rydberg coupling and interaction force}
\label{sec:rydberg-forces}
The transverse and longitudinal fields of the Ising Hamiltonian are generated via a two-photon transition that couples the ground state to a Rydberg state.
In particular, as explained in the main text, the laser detuning controls the longitudinal field $h_z = \Delta-\widetilde{\mathcal{N}}_c U_0/2$, while the Rabi frequency sets the transverse field $h_x = \Omega$.
The choice of Rydberg state and the interatomic distance define the interaction energy scale $U_0$, which corresponds to the van der Waals interaction.

We use two Ti:Sa lasers for the two-photon transition, with a \qty{5}{\gauss} field applied along the propagation axis.
The first laser is at \qty{810}{\nano\meter} and, after a doubling cavity, reaches \qty{405}{\nano\meter}, which is used to couple the ground state $\ket{4S_{1/2}, F=2, m_F=2}$ to the intermediate state $\ket{5P_{3/2}, F=3, m_F=3}$.
The second laser is at \qty{974}{\nano\meter}, which we then frequency-tune to address the two different Rydberg states used in the experiment.
Both lasers are frequency-locked to a ULE cavity via a Pound-Drever-Hall scheme.
The two Rydberg beams are elliptically shaped to couple the full array as uniformly as possible; the vertical (horizontal) waists are \qty{35}{\micro\meter} (\qty{300}{\micro\meter}) for the \qty{405}{\nano\meter} beam and \qty{10}{\micro\meter} (\qty{170}{\micro\meter}) for the \qty{974}{\nano\meter} beam.
We set the intermediate-state detuning to \qty{500}{\mega\hertz} and achieve a maximum Rabi frequency of roughly $\Omega_{\text{max}} = 2\pi \times \qty{3.5}{\mega\hertz}$.
We estimate a maximum \qty{10}{\percent} difference in Rabi frequency between edge atoms and central atoms.
Both lasers are intensity-stabilized using an acousto-optic modulator (AOM) double-pass with photodiodes, in a sample-and-hold scheme.
The AOMs are also used for frequency tuning and to control the timing and duration of the pulses.
The switching time is limited by the \qty{405}{\nano\meter} double-pass, for which we measured a rise time of \qty{200}{\nano\second}.

\begin{figure}[t]
	\centering \includegraphics{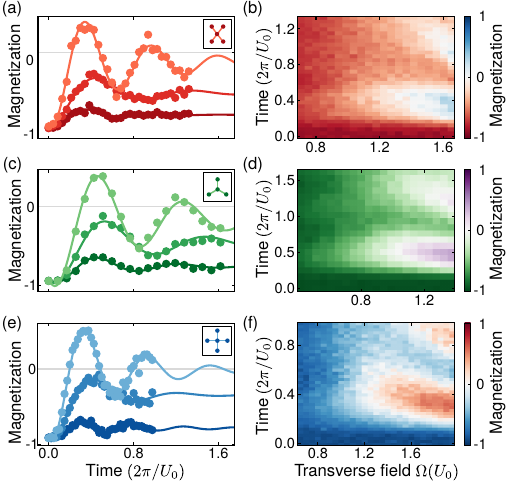}
	\caption{
		Post-quench magnetization dynamics of $M_z$ in the kagome and honeycomb lattices and the large $15 \times 15$ square array, row-wise as identified by the pictogram-insets in (a) and (c).
		\textbf{(a)}, \textbf{(b)} Kagome lattice.
		The traces in (a) are for $\Omega/U_0 = \numlist{0.7;1.1;1.6}$.
		\textbf{(c)}, \textbf{(d)} Honeycomb lattice.
		The traces in (c) are for $\Omega/U_0 = \numlist{0.6;0.9;1.4}$.
		\textbf{(e)}, \textbf{(f)} $15 \times 15$ square lattice.
		The traces in (e) are for $\Omega/U_0 = \numlist{0.7;1.1;1.85}$.
		Statistical error bars of one standard error of the mean are smaller than the marker size.
	}
	\label{fig:si1}
\end{figure}

The experimental sequence proceeds as follows: once the atoms are initialized in the fully magnetized state, we adiabatically ramp down the tweezer power to a  trap depth of $U_0/\hbar=2\pi \times \qty{2}{\mega\hertz}$, then switch the tweezers off entirely and apply the pulse.
Immediately after the pulse, we switch the tweezers back on: atoms in the ground state remain trapped, while atoms in the Rydberg state are expelled from the tweezers due to the ponderomotive potential.
By measuring the recapture probability, we determine the probability of the atom being in the ground state (spin down).
During this procedure we typically lose \qty{5}{\percent} of the atoms, which results in an offset of our data that does not affect any of the conclusions.

The van der Waals potential is characterized by a strong gradient $\partial V/\partial d = -6V/d$, resulting in a strong repulsive force.
This spin-motion coupling leads to a time-dependent interaction strength and finally to decoherence.
In our case, we can safely neglect this effect on the timescales we probe.
If we consider the worst-case scenario of a pair of atoms with $U_0 = 2\pi \times \qty{1.7}{\mega\hertz}$, $d_0 = \qty{7}{\micro\meter}$, and interaction time $t_{\text{int}} = \qty{1.2}{\micro\second}$, the change in interaction strength over that time is only $\sim\qty{2}{\percent}$.

\begin{figure}[t]
	\centering \includegraphics{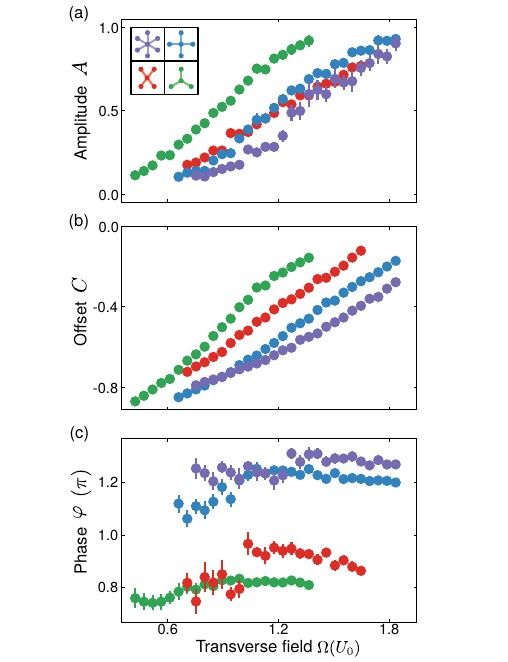}
	\caption{
		Fit results for \textbf{(a)} amplitude $A$, \textbf{(b)} offset $C$, and \textbf{(c)} phase $\varphi$ across the four lattice geometries, as a function of the transverse field.
		The phase changes quickly around the observed frequency minimum.
	}
	\label{fig:si2}
\end{figure}

\subsection{Magnetization dynamics data and fit results}
\label{sec:magnetization-data}
Here, we present the magnetization data for the arrays not shown in the main text, along with the remaining fit parameters.
The fit function used for the magnetization data is $M_z(t) = A\,e^{-\gamma t}\cos(\omega t + \varphi) + C$.
\autoref{fig:si1} shows the magnetization dynamics for the kagome, honeycomb, and square $15\times15$ arrays.
The total pulse time is typically $t_{\max} \sim \qty{1}{\micro\second}$, with approximately 40 experimental shots per time and transverse field.

\autoref{fig:si2} shows the fitted amplitude $A$, offset $C$, and phase $\varphi$ as a function of the transverse field for the different arrays.
At fixed transverse field, the oscillation amplitude decreases with increasing coordination number, consistent with MF predictions, since the critical point shifts to higher values as the coordination number increases.
The offset follows the same trend.
The phase, on the other hand, appears to transition between two values in the vicinity of the critical point, taking on two distinct values in the interaction-dominated regime and the transverse-field-dominated one.

\subsection{Mean-field description and frequency locking}
\label{sec:mf}
Within the mean-field (MF) approximation, the two-dimensional transverse-field Ising model of \autoref{eq.hamiltonian} reduces to a single collective spin precessing around a magnetic field.
For vanishing longitudinal field, $h_z=0$, the equations of motion read
\begin{equation}
	\dot{\vec{M}} = \vec{M} \times \vec{B}, \qquad \vec{B} = \left(-\Omega,\, 0,\, -\frac{U_0 \widetilde{\mathcal{N}}_c}{2} M_z\right), \label{eq.motion}
\end{equation}
where $\vec{M} = (M_x,\, M_y,\, M_z)$ denotes the magnetization per site averaged over the lattice.
These equations conserve both the MF energy,
\begin{equation}
	E_{\mathrm{MF}} = \frac{U_0 \widetilde{\mathcal{N}}_c}{8} M_z^2 + \frac{\Omega}{2} M_x ,
\end{equation}
and the spin length $|\vec{M}|^2 = M_x^2 + M_y^2 + M_z^2$, which allows the dynamics to be recast as that of a one-dimensional particle with mass $m=1$ in an effective potential,
\begin{equation}
	V(M_z) = \frac{1}{2}\left(\Omega^2 - U_0 \widetilde{\mathcal{N}}_c E_0\right) M_z^2 + \frac{(U_0 \widetilde{\mathcal{N}}_c)^2}{32} M_z^4 ,
\end{equation}
where $E_0 = U_0 \widetilde{\mathcal{N}}_c / 8$ is the energy of the initial state before the quench, in our case the fully magnetized state with all spins down.
Since the quench energy is fixed, the ratio $\Omega/(U_0\widetilde{\mathcal{N}}_c)$ determines the shape of the potential, and two regimes can be distinguished.
For $\Omega \gg U_0 \widetilde{\mathcal{N}}_c$ the potential is harmonic, describing essentially non-interacting Rabi oscillations, whereas for $\Omega \lesssim U_0 \widetilde{\mathcal{N}}_c/(2\sqrt{2})$ the potential is characterized by a double-well structure.
In the latter regime, the dynamics depends crucially on the energy of the central maximum of the potential.
If the quench energy $E_0$ lies below the energy of the maximum, the magnetization oscillates within a single well.
When $E_0$ exactly matches it, the magnetization reaches $M_z = 0$ and comes to rest there.
This condition defines the critical value of the transverse field, $\Omega_{\mathrm{crit}} = U_0 \widetilde{\mathcal{N}}_c / 4$, quoted in the main text.
For $\Omega > \Omega_{\mathrm{crit}}$, the system explores the full range of the magnetization, oscillating between $M_z = -1$ and $M_z = +1$.
In summary, $\Omega_{\mathrm{crit}}$ marks the threshold beyond which the magnetization reverses.

\begin{figure}[t]
	\centering
	\includegraphics[width=0.98\linewidth]{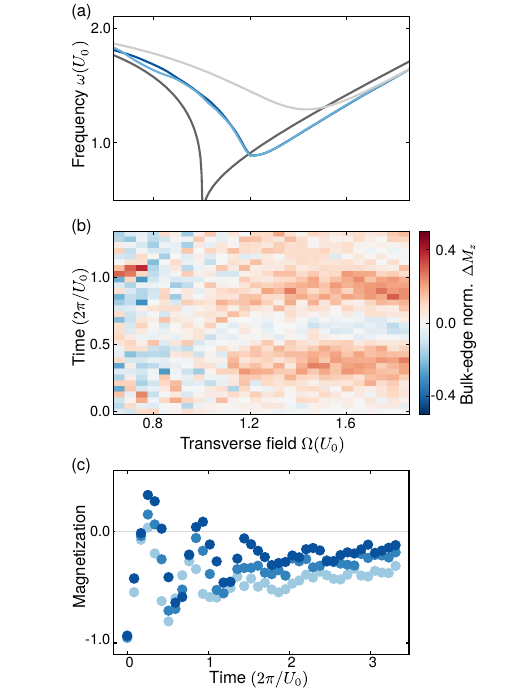}
	\caption{Mode softening in the two-population MF model and bulk, edge, and corner dynamics.
		\textbf{(a)} Mode softening in the two-population MF model: in gray, $\alpha=1$ (dark) and $\alpha=0$ (light), corresponding to the cases of only bulk or only edge spins, respectively.
		The coupled case with $\alpha = 0.4$ is shown in dark (light) blue for bulk (edge) spins.
		This value of $\alpha$ corresponds to the edge-to-bulk fraction estimated for our experiments.
		The broadening of the mode-softening feature is also apparent here.
		\textbf{(b)} Normalized difference $\Delta M_z$ between bulk and edge spins.
		\textbf{(c)} Site-resolved magnetization dynamics for a single bulk, edge, and corner site (from darker to lighter), at a transverse field value of $\Omega/U_0 = 1.2$.
	}
	\label{fig:si3}
\end{figure}

From this picture, one can already anticipate qualitatively how the oscillation frequency depends on the transverse field.
For $\Omega < \Omega_{\mathrm{crit}}$, the oscillation frequency decreases monotonically with increasing $\Omega$, vanishing at $\Omega_{\mathrm{crit}}$, where the dynamics comes to rest at $M_z = 0$.
Above $\Omega_{\mathrm{crit}}$, the system resumes oscillating behavior and the frequency grows with $\Omega$.
MF theory thus predicts a minimum of the oscillation frequency at the critical transverse field, the mode softening discussed in the main text.

In the MF model presented, one assumption is that all atoms share the same coordination number.
This is not the case in our experiment, because of finite-size effects and imperfect loading arising from the rearrangement procedure.
We can therefore distinguish two spin populations: one with the nominal coordination number and zero longitudinal field, and the other with one fewer nearest neighbor and a longitudinal field $h_z=U_0/2$.
If the two populations were independent, MF theory would predict strongly different behavior.
In our system, however, they are tightly coupled via the interactions, which results in a common oscillation frequency.
This effect can be explained with a simple extension of the MF model.
Consider that the array is populated by a fraction $\alpha$ of bulk and $(1-\alpha)$ of edge spins.
The magnetic fields seen by the two populations are
\begin{equation}
	\begin{aligned}
		\vec{B}_{\mathrm{bulk}} & = \left(-\Omega,\; 0,\; -\frac{U_0 \widetilde{\mathcal{N}}_c}{2} M_z\right),                     \\
		\vec{B}_{\mathrm{edge}} & = \left(-\Omega,\; 0,\; -\frac{U_0 (\widetilde{\mathcal{N}}_c-1)}{2} M_z + \frac{U_0}{2}\right),
	\end{aligned}
\end{equation}
where $\widetilde{\mathcal{N}}_c$ denotes the effective coordination number, which accounts for the van der Waals tail beyond nearest neighbors, and $M_z = \alpha M_{z,\mathrm{bulk}} + (1-\alpha) M_{z,\mathrm{edge}}$ is the total magnetization of the system, through which the edge and bulk spins are coupled.
In \autoref{fig:si3} we show the mode-softening curves for the two populations alone (i.e., the decoupled case, $\alpha=0$ or $1$) and for the coupled case ($\alpha=0.4$).
Despite the bulk and edge populations experiencing different effective fields, their coupling through the total magnetization $M_z$ locks their oscillation frequencies together, in contrast to the naive expectation of two independent softening curves for the decoupled case.
Also shown in \autoref{fig:si3} is the difference in magnetization dynamics between the average of all bulk sites and all edge sites.
We plot $\Delta M_z = (M_{z,\mathrm{bulk}}-M_{z,\mathrm{edge}})/(\max(M_{z,\mathrm{bulk}})-\min(M_{z,\mathrm{bulk}}))$ normalized to account for the bias that would otherwise arise from the oscillation amplitude alone.
In the interaction-dominated regime, no significant difference is observed.
In the transverse-field-dominated regime, however, the dynamics differ slightly, not in oscillation frequency but rather in the amplitude and phase of the oscillation.
This is also confirmed by the site-resolved dynamics for a single bulk, edge, and corner site at $\Omega/U_0=1.2$.

\subsection{Discrete truncated Wigner approximation}
\label{sec:DTWA}

In the main text, the experimental data were compared with semiclassical predictions obtained via the discrete truncated Wigner approximation. We refer the reader to the original Ref.~\cite{schachenmayer2015-jch} for the justification and scope of the method; here we report only the main steps needed to implement it.
\begin{enumerate}
	\item The Pauli operators at each site $\hat{X}_i,\hat{Y}_i,\hat{Z}_i$ are replaced with classical spins $s_i^x,s_i^y,s_i^z$.
	\item The initial \enquote{all-down} state is sampled from the discrete Wigner function, yielding either $s_i(0) = (1,-1,-1)$ or $s_i(0) = (-1,1,-1)$ independently for each site, with equal probability.
	\item The quantum evolution is replaced with the Hamiltonian evolution of a family of $N_\mathrm{traj}$ classical trajectories $\{\{s_{i,n}^{x,y,z}(t)\}_{i=1}^N\}_{n=1}^{N_\mathrm{traj}}$, each evolving independently under the dynamics generated by (the classical limit of) the quantum Hamiltonian in \autoref{eq.hamiltonian}. A second-order, symplectic Suzuki--Trotter algorithm is utilized, with $H$ decomposed into 4 sublattices~\cite{Omelyan2001Algorithm}: this is needed to incorporate the next-nearest-neighbor interactions of the underlying Rydberg Hamiltonian.
	\item The expectation values of observables are computed with the rules
	      \begin{equation}
		      \langle \hat{X}_i(t) \rangle \approx \frac{1}{N_\mathrm{traj}} \sum_{n=1}^{N_\mathrm{traj}} s_{i,n}^x(t),
	      \end{equation}
	      and similarly for the other spin operators, and combinations thereof (where commutation relations do not matter). Notice that the spins are not normalized to have length 1, but all expectation values of Pauli operators will take values between $-1$ and $+1$, after the trajectory average is taken.
\end{enumerate}

For the DTWA simulations, a $16\times16$ square lattice geometry was used, with no vacancies, and $N_\mathrm{traj}=40000$.
All the other parameters were fixed in accordance with the experiment.

\subsection{Tree-tensor-network simulations}
\label{sec:ttn}
For the $10\times10$ square lattice, we compare the dynamics observed in the Rydberg arrays with numerical simulations aimed at reproducing as realistically as possible the experimental setup. This includes finite next-nearest-neighbor interactions, spatial inhomogeneity of the transverse field, and rearrangement efficiency.

\begin{figure}[t]
	\centering
	\includegraphics[width=0.98\linewidth]{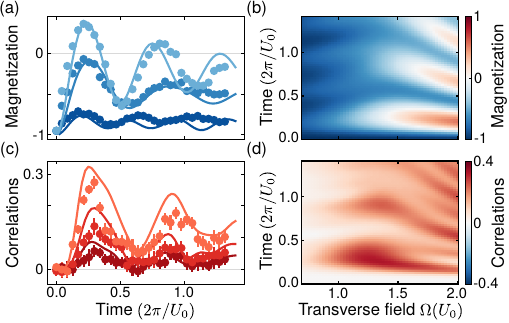}
	\caption{Comparison between TTN and experiment of the correlation and magnetization dynamics. \textbf{(a)} Magnetization dynamics of $M_z$ for $\Omega/U_0 = \numlist{0.75;1.35;1.8}$, from darker to lighter color. Solid lines are TTN simulations. \textbf{(b)} Magnetization dynamics TTN color plot. \textbf{(c)} Correlation dynamics of $C^{(1)}$ for $\Omega/U_0 = \numlist{0.75;0.95;1.35}$. Solid lines are TTN simulations. \textbf{(d)} Correlation dynamics TTN color plot.
	}
	\label{fig:si4}
\end{figure}

In the experiment, the filling fraction after rearrangement is $p\approx\qty{90}{\percent}$. To reproduce this, each site is randomly chosen to be filled with probability $p$. Both vacancies and open boundaries reduce a site's number of
interaction partners in the same way. Let $n_i=\sum_{j\in\mathrm{NN}(i)} m_j$ be the number of present
nearest neighbors of site $i$, where $m_i=0$ for empty sites and $1$ otherwise. A site with all $4$ neighbors present
has $n_i=4$, an edge site has at most $n_i=3$, and a corner at most $n_i=2$. Each missing neighbor reduces $n_i$ by one further. Analogously, $\tilde n_i=\sum_{j\in\mathrm{NNN}(i)} m_j$ counts present diagonal (next-nearest) neighbors out of a maximum of $4$. In the experiment, a constant detuning $\Delta$ is chosen to cancel the bulk longitudinal field in the ideal case where all four neighbors are present, i.e., $\Delta=2\,U_0$ as explained in the main text. As a result, at the edges or in the vicinity of vacancies the detuning introduces a large longitudinal field $(4-n_i)U_0/2$ of the same order of magnitude as the antiferromagnetic coupling. Next-nearest-neighbor interactions generate an additional longitudinal field $\tilde n_i U_0/8$ which is not compensated for in the experiment. Once the vacancy pattern is fixed, the nearest-neighbor interaction $U_0$ sets the value of all NNN interactions and local longitudinal fields. If it is fixed to unity (i.e., used as the reference energy scale), the only free parameter of the model is the transverse field $\Omega/U_0$. To reproduce the transverse field inhomogeneity present in the experiment, the transverse field at each site is weighted by the Gaussian envelope of the two Rydberg beams. This weighting varies by roughly \qty{10}{\percent} from the edge to the center of the array.

All the numerical simulations rely on the tensor-network library \texttt{qtealeaves} \cite{qtealeaves}. We represent quantum states as TTNs with a fixed maximum bond dimension $\chi=128$ throughout the time evolution. The initial state is the experiment's fully magnetized
$\ket{\downarrow \downarrow \downarrow \dots \downarrow \downarrow}$ state, which is then quenched with the Hamiltonian described above for different values of $\Omega$, and time evolved using the time-dependent variational principle (TDVP) with a fixed time step $dt=0.025$ up to $t_{\mathrm{max}}=1.4$ (in units of $2\pi/U_0$). The observables that we compute are the site-resolved magnetization $\langle \hat Z_i\rangle$ and the
nearest-neighbor connected correlator
$\langle \hat Z_i \hat Z_j\rangle-\langle \hat Z_i\rangle\langle \hat Z_j\rangle$. Spatial averages are always restricted to non-vacant sites.
\autoref{fig:4} of the main text includes TTN results for the variance of the magnetization $\overline{\mathrm{var}}(\hat M_z) = N( \langle \hat M_z^2\rangle - \langle \hat M_z\rangle^2)/ (1-\langle \hat M_z\rangle^2 )$. As $\hat M_z=\frac{1}{N}\sum_i \hat Z_i$, this requires computing all the correlators $C_{ij}=\langle \hat Z_i \hat Z_j\rangle$.

In~\autoref{fig:si4} we report the dependence of the observables on the transverse field $\Omega/U_0$. Every curve starts from the fully magnetized state with all spins down and shows a sharp initial excursion followed by partial revivals and a slow drift toward a
quasi-stationary plateau. In the left panels we overlay the TTN prediction directly on the
experimental data. The panels on the right show the scan as a 2D density plot.

\subsection{Semiclassical theory of damping} \label{sec:semiclassical}

The simplest starting point for quantitative analytics of potentially universal phenomena in a spin system is a semiclassical treatment.
Here, we perform such a large-$S$ analysis and find that it accounts for the central experimental findings, namely the collapse of appropriately rescaled curves of oscillation frequency and damping rate across lattices and strengths of initial fluctuations.
First, we rewrite the Hamiltonian as
\begin{equation}
	\label{eq:H_Ising_semiclassics}
	H = -\frac{1}{2} \sum_{ij} J_{ij} s_i^z s_j^z - \Gamma \sum_i s_i^x.
\end{equation}
The quantities $s_i^\alpha$ are normalized, unit-length spin operators, $s_i^\alpha = \hat{S}_i^\alpha/S$.
This is useful for treating the classical limit $S \to \infty$ and the (perturbative approaches to the) quantum case on the same footing.
Comparing to \autoref{eq.hamiltonian}, $J_{ij} \equiv U_{ij}/4$ and $\Gamma \equiv -\Omega/2$, having taken into account the low-to-high energy mapping.

The equations of motion for the spins are
\begin{equation}
	\dot{s}_i = \bigg( -\Gamma \hat{x} -\hat{z} \sum_{j} J_{ij} s_j^z \bigg) \times s_i.
\end{equation}

\subsubsection{Dynamics of the uniform mode}

The global magnetization $m^\alpha = \frac{1}{N}\sum_i s_i^\alpha$ has maximal overlap with the $k=0$ (uniform) mode of the Fourier transform of the spin configurations. If the system is initialized in a spatially homogeneous initial state, then the $k=0$ mode satisfies the same equation of motion as the collective spin in mean field in \autoref{sec:mf}:
\begin{equation}
	\label{eq:EoM_zero_mode}
	\dot{m} = (-\Gamma \hat{x} - \hat{z} m^z \tilde{J}_0) \times m \equiv -h_0 \times m,
\end{equation}
where $\tilde{J}_0 = \sum_j J_{ij}$ is the effective exchange energy per site. Following the same logic as \autoref{sec:mf}, the $k=0$ mode oscillates forever, with an oscillation frequency (in units of $\tilde{J}_0$) that is a universal function of $\Gamma/\tilde{J}_0$.

\subsubsection{Equation of motion for the fluctuations}

We now consider the effects of small spatial inhomogeneities atop the initial state at the linearized level.
Defining
\begin{equation}
	\delta s_i^\alpha = s_i^\alpha - m^\alpha,
\end{equation}
the linearized equations of motion read
\begin{equation}
	\label{eq:EoM_deltas}
	\delta \dot{s}_i \simeq \bigg( -\Gamma \hat{x} -\hat{z} m^z \tilde{J}_0 \bigg) \times \delta s_i -  \sum_{j} J_{ij} \delta s_j^z \hat{z} \times m.
\end{equation}
This is the only approximation until \autoref{eq:Hill}.
In Fourier space, one obtains
\begin{equation}
	\label{eq:EoM_deltas_k}
	\partial_t \delta \tilde{s}_k \simeq \big( -\Gamma \hat{x} -\hat{z} m^z \tilde{J}_0 \big) \times \delta \tilde{s}_k - \tilde{J}_k \delta \tilde{s}_k^z \hat{z}\times m.
\end{equation}

We now perform a straightforward computation to reduce \autoref{eq:EoM_deltas_k} to a more amenable form.
This computation was supported by Claude Fable 5.1.
The results have been carefully checked and the presentation has been extensively edited for clarity.
First, note that $m \perp \delta \tilde{s}_k$. Therefore,
\begin{equation}
	\label{eq:orthonormal_decomp}
	\delta \tilde{s}_k = \eta_{1,k} e_1 + \eta_{2,k} e_2,
\end{equation}
where
\begin{align}
	e_1 & = \frac{1}{q}\, m \times \hat z, \qquad q = \| m \times \hat z \| = \sqrt{1-(m^z)^2}, \\
	e_2 & = m \times e_1.
\end{align}

As the linearized theory decouples the Fourier modes, we suppress the $k$-dependence in $\eta$ to ease the notation.
Plugging \autoref{eq:orthonormal_decomp} into \autoref{eq:EoM_deltas_k}, and then projecting on $e_1$ and $e_2$, one finds
\begin{equation}
	\dot{\eta}_1 - \eta_2 (\dot{e}_1 \cdot e_2 + h_0 \cdot m) - \tilde{J}_k \delta \tilde{s}_k^z q = 0
\end{equation}
and
\begin{equation}
	\dot{\eta}_2 + \eta_1 (\dot{e}_1 \cdot e_2 + h_0 \cdot m) = 0.
\end{equation}
respectively.
The first equation is further simplified by using
\begin{align}
	\delta \tilde{s}_k^z
	 & = \hat{z} \cdot (\eta_1 e_1 + \eta_2 e_2)
	= \eta_2 \, \hat{z} \cdot ( m \times e_1 ) \nonumber \\
	 & = \eta_2 \, e_1 \cdot (\hat{z} \times m)
	= - \eta_2 q \, e_1 \cdot e_1
	= -\eta_2 q,
\end{align}
thus finding
\begin{equation}
	\dot{\eta}_1 - \eta_2 (\dot{e}_1 \cdot e_2 + h_0 \cdot m - \tilde{J}_k q^2) = 0.
\end{equation}

The equations for $\eta_1$ and $\eta_2$ reduce to a simple form as $\dot{e}_1 \cdot e_2 + h_0 \cdot m$ is a constant in time. A possible way to see it is by expressing everything in coordinates. Introduce
\begin{align}
	m   & = (q \cos \phi, q \sin \phi, m^z),    \\
	e_1 & = (\sin \phi, -\cos \phi, 0)          \\
	e_2 & = (m^z \cos \phi, m^z \sin \phi, -q), \\
	h_0 & = (\Gamma,0, \tilde{J}_0 m^z).
\end{align}
Direct computation leads to
\begin{equation}
	\label{eq:temp1}
	\dot{e}_1 \cdot e_2 + h_0 \cdot m = \Gamma q \cos \phi + \tilde{J}_0 (m^z)^2 + \dot{\phi} m^z.
\end{equation}
The quantity $\dot \phi$ can be found by computing
\begin{equation}
	\dot{m}^x m^y - m^x \dot{m}^y = - q^2 \dot{\phi}
\end{equation}
from the coordinate expression, and
\begin{equation}
	\dot{m}^x m^y - m^x \dot{m}^y = \tilde{J}_0 m^z [(m^x)^2 +(m^y)^2] - \Gamma m^x m^z
\end{equation}
from the equations of motion, \autoref{eq:EoM_zero_mode}.
Comparing the two, one fixes
\begin{equation}
	\label{eq:phi_dot}
	\dot \phi = \frac{\Gamma m^x m^z}{1-(m^z)^2} -\tilde{J}_0 m^z.
\end{equation}

The last step is to realize that the quantity
\begin{equation}
	-\Gamma q \cos \phi - \frac{1}{2} \tilde{J}_0 (m^z)^2 = \frac{E_{\mathrm{MF}}}{N}
\end{equation}
is the mean-field energy density.
It can be replaced with its value at $t=0$ when the state (to this order) is the fully polarized state $m^z=-1$:
\begin{equation}
	\label{eq:MF_energy_relation}
	-\Gamma q \cos \phi - \frac{1}{2} \tilde{J}_0 (m^z)^2 = -\frac{1}{2} \tilde{J}_0.
\end{equation}

Plugging back Eqs.~\ref{eq:phi_dot} and \ref{eq:MF_energy_relation} into \autoref{eq:temp1}, and simplifying algebraic factors, one finally finds
\begin{equation}
	\dot{e}_1 \cdot e_2 + h_0 \cdot m = \frac{1}{2}\tilde{J}_0.
\end{equation}

Now the equations of motion for $\eta_1$ and $\eta_2$ read
\begin{equation}
	\dot{\eta}_1 - \eta_2 \left( \frac{1}{2}\tilde{J}_0 - \tilde{J}_k q^2 \right) = 0, \qquad
	\dot{\eta}_2 + \frac{\tilde{J}_0}{2}\eta_1 = 0.
\end{equation}
Combining them yields
\begin{equation}
	\label{eq:Hill}
	\ddot{\eta}_{2,k}(t) + \frac{\tilde{J}_0}{2} \left( \frac{\tilde{J}_0}{2} - \tilde{J}_k \big[1-m^z(t)^2\big]\right) \eta_{2,k}(t) = 0,
\end{equation}
where we have reinstated the explicit $k$ dependence on $\eta$.

\subsubsection{Growth of fluctuations}

Equation~\ref{eq:Hill} predicts that the fluctuations around the uniform mode grow as a parametric instability of a periodically forced system.
While the periodic forcing is in origin self-induced, to this level of approximation (linear spin-wave theory) there is no feedback of the fluctuations onto the uniform mode.

The exponential growth of fluctuations implies a decay of the uniform mode,
as follows from the ``sum rule''
\begin{equation}
	m^z(t)^2 \leq 1 - \frac{1}{N} \sum_k \tilde{C}_c(k;t),
\end{equation}
where $\tilde{C}_c(k;t)$ is the Fourier transform of
\begin{equation}
	\label{eq:corrfunc}
	C_\mathrm{c}(r;t) = \langle s_i^z(t) s_{i+r}^z(t) \rangle_\mathrm{is} - \langle s_i^z(t) \rangle_\mathrm{is} \langle s_{i+r}^z(t) \rangle_\mathrm{is}.
\end{equation}
The averages are over a family of initial states `is', which are assumed to be translationally invariant. To prove the sum rule, just notice that
\begin{align}
	\frac{1}{N} \sum_k \tilde{C}_c(k;t) & = C_\mathrm{c}(0;t)
	= \langle s_i^z(t)^2 \rangle_\mathrm{is} - \langle s_i^z(t) \rangle_\mathrm{is}^2 \nonumber \\
	                                    & = \langle s_i^z(t)^2 \rangle_\mathrm{is} - m^z(t)^2
	\leq 1 - m^z(t)^2.
\end{align}

\begin{figure*}
	\centering
	\includegraphics{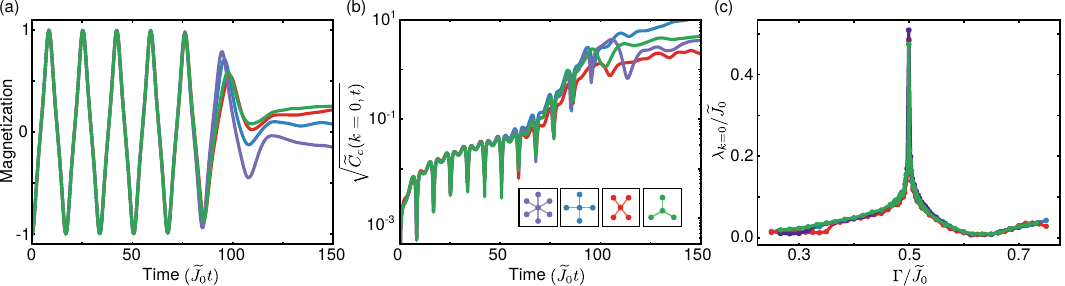}
	\caption{
		Simulation results for the semiclassical model in \autoref{eq:H_Ising_semiclassics}.
		\textbf{(a)} The uniform mode $m^z(t)$ oscillates in time in all lattice geometries; the curves lie on top of each other at early times when the $x$-axis is rescaled by the coordination-number-dependent factor $\tilde{J}_0$.
		The oscillations stop at about the same rescaled time.
		The inverse of this time-scale is a measure of the damping rate.
		\textbf{(b)} The fluctuations, as measured by the correlation function in \autoref{eq:corrfunc} with $k=0$, grow exponentially in time at the same rate in all lattices.
		Once the fluctuations are comparable to the mean $m^z(t)$, the growth stops and lattice effects are visible.
		The timescale when the fluctuations become comparable to the mean $m^z(t)$ is the same as the timescale when the oscillations of $m^z(t)$ shown in (a) stop.
		The data shown in (a) and (b) are for $\Gamma = 0.55 \tilde{J}_0 > \Gamma_\mathrm{crit}$.
		\textbf{(c)} The growth rates $\lambda_k$ extracted from fits $\tilde{C}_\mathrm{c}(k;t) \simeq A_k e^{2 \lambda_k t}$, for the $k=0$ mode.
		The peak appears precisely at the mode softening transition. 
	}
	\label{fig:semiclassics}
\end{figure*}

The timescale of the two processes---the decay of $m^z(t)$ and the growth of fluctuations---appear roughly the same; this allows one to connect the damping rate $\gamma$ of $m^z(t)$ with the growth exponent $\lambda$ of the increase $\eta_{2,k}(t) \sim e^{\lambda_k t} \eta_{2,k}(0)$.

Equation~\ref{eq:Hill} is a Hill equation, and is in principle amenable to analytical solution for some particular choices of parameters---even using the exact analytic expression of $m^z(t)$, which involves elliptic functions.
Here, we use it to argue why there should be a collapse in the rates $\lambda$ describing the \emph{overall} growth of fluctuations.
The main point is that the dependence on the lattice geometry is mediated only by $\tilde{J}_k$.
Assuming that the uniform mode is dominantly destabilized by the $k$-mode that grows the most (rather than, say, a simple \enquote{average over the Brillouin zone}), the lattice role is subdominant to the values of the parameter $\tilde{J}_0$---which is at the heart of the collapse---and the frequency of the forcing term $\omega = \tilde{J}_0 f(\Gamma/\tilde{J}_0)$, each of which obeys the scaling collapse.

The computation above is linear in the fluctuations around the mean-field.
It encompasses both the case of classical rotors of norm $1$, and of quantum spins-$S$.
The degree of accuracy is of course dependent on the size of the representation $S$, with $S=1/2$ being the least controlled case.
What changes is the initial condition $s_i^\alpha(0)$, which would be thermal noise for the classical case, and quantum noise, e.g.\ described by the discrete truncated Wigner scheme, in the quantum case.
The fact that such a semiclassical prediction captures the behavior of a spin-$1/2$ system constitutes a second important finding of this work.

The fact that fluctuations grow exponentially can be read off \autoref{eq:Hill} directly.
When
\begin{equation}
	\frac{\tilde{J}_0}{2} - \tilde{J}_k \big[1-m^z(t)^2\big] < 0,
\end{equation}
the solution is given (locally in time) by two exponentials, one of which is growing.
The rate is maximal at the minima of $m^z(t)^2$, i.e.\ the zero-crossings of the magnetization.
The condition is satisfied for $\Gamma$ close enough to the dynamical phase transition, where $m^z(t)$ either reaches close to zero or crosses it altogether, switching sign, and for $k$ close enough to 0.
It is therefore the long-wavelength modes that are the most unstable.

We benchmark the perturbative description of the growth of fluctuations against direct numerical simulations of the semiclassical equations of motion, for all lattice geometries. The parameters are fixed as in the main text---except that a larger system size of 50$\times$50 spins is utilized, with periodic boundary conditions. The results are shown in Fig.~\ref{fig:semiclassics}. To measure the strength of the fluctuations, we employ the correlation function $\tilde{C}_\mathrm{c}(k;t)$, see Eq.~\eqref{eq:corrfunc}. The family of initial states `is' is such that all the states are characterized by the same amount of noise: each spin is perturbed independently at time $t=0$, isotropically about $s^z = -1$, by a polar angle $\theta \sim |\text{normal}(0,\epsilon)|$. For the simulations, $\epsilon = 0.003$.

\end{document}